\documentclass[notitlepage,12pt]{article}
\usepackage[left=1.8cm,right=1.8cm,top=1.8cm,bottom=1.5cm]{geometry}
\usepackage[fleqn]{amsmath}
\usepackage{amssymb,amsthm}
\usepackage{bm}
\usepackage{setspace}
\usepackage{subfig}
\usepackage[pdftex,colorlinks,citecolor = black,linkcolor = black]{hyperref}%
\usepackage{xcolor}
\usepackage{authblk}
\usepackage[authoryear]{natbib}
\usepackage{graphicx}
\allowdisplaybreaks

\usepackage{etoolbox}

\usepackage{algorithmic}
\usepackage[plain]{algorithm}
\numberwithin{algorithm}{section}
\floatstyle{plaintop}
\restylefloat{algorithm}
\usepackage{caption}
\usepackage[normalem]{ulem} 
\usepackage{comment}

\BeforeBeginEnvironment{equation}{\begin{singlespace}\setlength{\abovedisplayskip}{-2.5ex}\noindent\ignorespaces}
\AfterEndEnvironment{equation}{\end{singlespace}\setlength{\belowdisplayskip}{-3ex}\noindent\ignorespaces}
\BeforeBeginEnvironment{equation*}{\begin{singlespace}\setlength{\abovedisplayskip}{-3ex}\noindent\ignorespaces}
\AfterEndEnvironment{equation*}{\end{singlespace}\setlength{\belowdisplayskip}{-3ex}\noindent\ignorespaces}
\BeforeBeginEnvironment{subequations*}{\begin{singlespace}\setlength{\abovedisplayskip}{-3ex}\noindent\ignorespaces}
\AfterEndEnvironment{subequations*}{\end{singlespace}\setlength{\belowdisplayskip}{-3ex}\noindent\ignorespaces}
\BeforeBeginEnvironment{subequations}{\begin{singlespace}\setlength{\abovedisplayskip}{-7ex}\noindent\ignorespaces}
\AfterEndEnvironment{subequations}{\end{singlespace}\setlength{\belowdisplayskip}{-3ex}\noindent\ignorespaces}

\BeforeBeginEnvironment{align}{\begin{singlespace}\noindent\ignorespaces}
\AfterEndEnvironment{align}{\end{singlespace}\noindent\ignorespaces}
\BeforeBeginEnvironment{align*}{\begin{singlespace}\noindent\ignorespaces}
\AfterEndEnvironment{align*}{\end{singlespace}\noindent\ignorespaces}

\numberwithin{equation}{section}

\newcommand{\appsection}[1]{
\let\oldthesection\thesection
  \renewcommand{\thesection}{Appendix \oldthesection}
  \section{#1}\let\thesection\oldthesection }

\newcommand{\supplementarysection}[1]{
\let\oldthesection\thesection
  \renewcommand{\thesection}{Supplementary material \oldthesection}
  \section{#1}
	\let\thesection\oldthesection }

\newcommand{\supplementarysectionN}[1]{
\let\oldthesection\thesection
  \renewcommand{\thesection}{Supplementary material:}
  \section{#1}
	\let\thesection\oldthesection }

\newcommand{\refeq}[1]{(\ref{eq:#1})}
\newcommand{\refsec}[1]{\ref{sec:#1}}

\newcommand{\reftab}[1]{\ref{tab:#1}}
\newcommand{\refalg}[1]{\ref{Ta:#1}}
\newcommand{\Rset}{\mathbb{R}}

\newenvironment{keywords}{
  \noindent\textbf{Keywords:}}{
  \medskip}
	
\renewcommand{\thefootnote}{\fnsymbol{footnote}}

\title{A numerical method for the estimation of time-varying parameter models in large dimensions}
\author[1]{Stella Hadjiantoni}
\author[2,3]{Erricos Kontoghiorghes}
\affil[1]{University of Kent, UK}
\affil[2]{Cyprus University of Technology}
\affil[3]{Birkbeck University of London, UK}
\date{}

\begin{document}
	\maketitle

{\let\thefootnote\relax\footnote{Corresponding author: S. Hadjiantoni, School of Mathematics, Statistics \& Actuarial Science, University of Kent, Canterbury, Kent CT2 7FS, UK. Email address: s.hadjiantoni@kent.ac.uk}}%

\vspace*{-2.75em}
\begin{abstract}
\vspace*{-.5em}
\singlespacing
\noindent A novel numerical method for the estimation of large time-varying parameter (TVP) models is proposed. The updating and smoothing estimates of the TVP model are derived within the context of generalised linear least squares and through numerically stable orthogonal transformations. The method developed is based on computationally efficient strategies. The computational cost is reduced by exploiting the special sparse structure of the TVP model and by utilising previous computations. The proposed method is also extended to the rolling window estimation of the TVP model. Experimental results show the effectiveness of the new updating, window and smoothing strategies in high dimensions when a large number of covariates and regressions are included in the TVP model. 
\end{abstract}


\section{Introduction}\label{sec:Intro}

\doublespacing

\indent\par
The assumption that the coefficients of a linear model are constant over time is often invalid. Recently, models with time-varying structures have been adopted to explain inflation dynamics, to forecast macroeconomic variables under structural change and to model interest rates (Cogley and Sargent, 2005; Primiceri, 2005; Stock and Watson, 2009; Koop and Korobilis, 2013; Zhang and Wu, 2015). A model with time varying coefficients can be given a state space formulation. The most common approach is to use the Kalman filter to provide the updated values of the coefficients, as each new observation is acquired. The Kalman filter is a fast recursive method, especially in small dimensions, but it does not have good numerical properties \citep{paige:77}. It is based on matrix inverses, which can be ill-conditioned, and may be the reason for inaccurate results \citep{golub:65}. Specifically in a recursive method like the Kalman filter, where at every new data point estimates of the unknown parameters are obtained using previous computations, a numerical error at one iteration of the algorithm can be propagated through to future computations and produce inaccurate results \citep{higham:02}. Generalised least squares (GLS) have been applied on a univariate time-varying model in order to derive the Kalman filter and Kalman smoother estimators \citep{sant:77}. However, this approach is difficult to implement in practical problems as it requires the inversion of a large variance-covariance matrix which is computationally demanding and numerically unstable.


The contribution, herein, is to develop a novel numerical method for the estimation of the multivariate time-varying parameter (TVP) regression model. The proposed method estimates the TVP model by solving a generalised linear least squares problem which yields the best linear unbiased estimator of the model \citep{paige:79,kourouklis:81}. The updating estimates, when new data are acquired, and the smoothing estimates, when existing data are revised, are derived. The method is next extended to the window estimation of the model where data are added and deleted simultaneously. Numerical strategies which update the model to include the effect of new observations and which downdate the model to exclude the effect of old or obsolete observations are employed \citep{paige:78,hadjiantoni:16,hadjiantoni:18}. 
The novel method is reliable in delivering accurate estimation results, and computationally efficient which makes it feasible to estimate large TVP models. 
This is achieved in two ways. Firstly, by employing efficiently previous computations when new observations are acquired and by exploiting the sparsity of the multivariate TVP model \citep{gill:74,paige:79,golub:13}. Secondly, the computational tools which are mainly orthogonal transformations, have the property of being numerically stable and also the capability to limit the computational expense of the estimation procedures \citep{gill:74,paige:78,bjorck:96,golub:13}. Furthermore, the proposed algorithm 
does not require non-singular variance-covariance matrices and also postpones the inversion of matrices up to the last step. 

The paper is organised as follows. Section~\refsec{UpdateTVPmodel} introduces a new numerical method for the estimation of the TVP model based on orthogonal transformations. Section~\refsec{RecursiveTVPSUR} considers the multivariate TVP model where the regressions are contemporaneously correlated. The numerical estimation of the model is presented when observations are added to and/or removed from the model. The smoothing estimates of the model are also derived. Section~\refsec{Experiments} presents the computational results and finally, Section~\refsec{Conclusions} concludes.

\section{Numerical Estimation of the TVP Model}\label{sec:UpdateTVPmodel}

Consider the univariate time-varying parameter (TVP) model which is given by%
\begin{subequations}%
\begin{equation}%
\psi_t = \bm{x}_t \bm{\beta}_t + \epsilon_t, \quad \epsilon_t \sim ( 0, \sigma^2), \ \ t=1,\dots,M \\
\end{equation}%
\text{and}
\begin{equation}%
\bm{\beta}_t = \bm{\beta}_{t-1} + \bm{\eta}_t, \quad \bm{\eta}_t \sim ( \bm{0}, \sigma^2 \bm{\Sigma}_{\eta} ), \ \ t=1,\dots,M.
\label{eq:TVPmodeltB}%
\end{equation}%
\label{eq:TVPmodelt}%
\end{subequations}%
\noindent Here $\psi_t$ is the observation of the dependent variable $\bm{y}$ at time $t$, $\bm{x}_t\in\Rset^k$ is the row vector of explanatory variables\index{explanatory variables} at time $t$, $\bm{\beta}_t \in \Rset^{k}$ is the vector of the unknown coefficients which are evolving over time according to the random walk in \refeq{TVPmodeltB}, and $\epsilon_t$ and $\bm{\eta}_t$ are the error terms with zero mean and variance $\sigma^2$ and $\sigma^2 \bm{\Sigma}_{\eta}$, respectively \citep{cooley:73,cooley:76}. Also $\mathbb{E} (\epsilon_t \epsilon_{t'}) = 0$ if $t\neq t'$ and $\mathbb{E} ( \bm{\eta}_t \bm{\eta}^T_{t'} ) = \bm{0}$ if $t \neq t'$, $t=1,\dots,M$, where $M$ is the sample size. In addition, $\bm{\Sigma}_{\eta}$ is a known, symmetric and non-negative dispersion matrix. Furthermore, from \refeq{TVPmodelt}, it is easy to derive (see \cite{sant:77}) 
\begin{equation*}
\bm{\beta}_t = \bm{\beta}_{t-1} + \bm{\eta}_t =  \dots= \bm{\beta}_1 + \sum_{s=1}^t \bm{\eta}_s . 
\end{equation*}
Therefore the TVP model \refeq{TVPmodelt} up to time $t$ takes the following form
\begin{equation*}
\begin{pmatrix}
\psi_1 \\
\vdots \\
\psi_{t-1} \\
\psi_t
\end{pmatrix} = \begin{pmatrix}
										\bm{x}_1 \\
										\vdots \\
										\bm{x}_{t-1} \\
										\bm{x}_t
								\end{pmatrix} \bm{\beta}_t + \begin{pmatrix}
																								\epsilon_1 \\
																								\vdots \\
																								\epsilon_{t-1} \\
																								\epsilon_t
																				\end{pmatrix} - \begin{pmatrix}
																														\bm{x}_1 & \cdots & \bm{x}_1 \\
																														\vdots	 & \ddots & \vdots \\
																														\bm{0}	       & \cdots       & \bm{x}_{t-1} \\
																														\bm{0}		     & \cdots & \bm{0}  
																												\end{pmatrix} \begin{pmatrix}
																																					\bm{\eta}_2 \\
																																					\vdots \\
																																					\bm{\eta}_t
																																			\end{pmatrix}			
\end{equation*}
or conformably the compact form
\begin{equation}
\bm{y}_t = \bm{X}_t \bm{\beta}_t + \bm{e}_t - \bm{A}_t \bm{u}_t, \quad \bm{e}_t-\bm{A}_t \bm{u}_t \sim \left( \bm{0}, \sigma^2 \bm{\Omega}_t \right),
\label{eq:TVPmodelCompact}
\end{equation}
where $\bm{y}_t$ is the vector of observations for the dependent variable up to time $t$, $\bm{X}_t$, $\bm{e}_t$, $\bm{A}_t $ and $\bm{u}_t$ are analogously defined and  $\bm{\Omega}_t = \bm{I}_t + \bm{A}_t (\bm{I}_{t-1} \otimes \bm{\Sigma}_{\eta} ) \bm{A}^T_t$ \citep{sant:77}. The GLS estimator of the latter model is given by 
\begin{equation*}
\bm{\hat{\beta}}_{t} = ( \bm{X}_t^T \bm{\Omega}_t^{-1} \bm{X}_t)^{-1} \bm{X}_t^T \bm{\Omega}_t^{-1} \bm{y}_t.
\end{equation*}
However, the derivation of the GLS estimator is computationally costly and numerically unstable when $\bm{\Omega}_t$ is ill-conditioned \citep{paige:78,kourouklis:81}.  

An alternative procedure to the GLS methodology is to consider solving a generalised linear least squares problem\index{generalised linear least squares problem} (GLLSP), that is,
\begin{equation}
\underset{\bm{\beta}_t, \bm{v}_t}{\text{argmin}} \ \ \left\| \bm{v}_t \right\|^2 \ \ \text{subject to} \ \ \bm{y}_t = \bm{X}_t \bm{\beta}_t + \bm{C}_t \bm{v}_t,
\label{eq:GLLSPt1}
\end{equation}
where $\bm{C}_t \in \Rset^{t \times t}$ is upper triangular and non-singular such that $\bm{\Omega}_t = \bm{C}_t \bm{C}_t^T$, $\bm{v}_t$ is an arbitrary vector, $\bm{v}_t \sim (\bm{0}, \sigma^2 \bm{I}_t) $ and $ \left\| \cdot \right\| $ denotes the Euclidean norm. Observe that $\bm{\Omega}_t$ is not formed explicitly but instead its special structure is taken into account. That is, 
\begin{align*}
\begin{split}
\bm{\Omega}_t &= \bm{I}_t + \bm{A}_t (\bm{I}_{t-1} \otimes \bm{\Sigma}_{\eta} ) \bm{A}^T_t \\
              &= \begin{pmatrix}
										\bm{I}_t & \bm{A}_t (\bm{I}_{t-1} \otimes \bm{C}_{\eta} )
								 \end{pmatrix} \begin{pmatrix}
																	\bm{I}_t & \bm{A}_t (\bm{I}_{t-1} \otimes \bm{C}_{\eta} )
															 \end{pmatrix}^T,
\end{split}
\end{align*}
where $\bm{C}_{\eta}$ is the Cholesky factor of $\bm{\Sigma}_{\eta}$, i.e. \( \bm{\Sigma}_{\eta} = \bm{C}_{\eta} \bm{C}_{\eta}^T \). Then the RQ decomposition (RQD) of \( (	\bm{I}_t \ \ \bm{A}_t (\bm{I}_{t-1} \otimes \bm{C}_{\eta} ) ) \) gives
\begin{equation}														
\begin{pmatrix}
		\bm{I}_t & \bm{A}_t (\bm{I}_{t-1} \otimes \bm{C}_{\eta} )
\end{pmatrix} \bm{P}_{t,1}  = \begin{pmatrix} 
																	\bm{0} & \bm{C}_t
															\end{pmatrix}, 
\label{eq:RQD1}
\end{equation}
where \( \bm{P}_{t,1} \in \Rset^{((t-1)k) \times ((t-1)k)} \) is orthogonal and \( \bm{C}_t \in \Rset^{t \times t} \) is upper triangular and non-singular. To solve \refeq{GLLSPt1}, the generalised QR decomposition\index{decomposition!generalised QR decomposition} (GQRD) of $\bm{X}_t$ and $\bm{C}_t$ is computed, namely,
\begin{subequations}
\begin{equation}
\bm{Q}^T_t \begin{pmatrix} 
							\bm{X}_t & \bm{y}_t 
					 \end{pmatrix} = \begin{pmatrix}
															\bm{R}_t & \bm{y}_{t,A} \\
															\bm{0}   & \bm{y}_{t,B}
													\end{pmatrix} \begin{matrix} k \hfill \\ t-k \end{matrix}
\end{equation}
\text{and}
\begin{equation}
\left( \bm{Q}^T_t \bm{C}_t \right) \bm{P}_{t,2} = \bm{U}_t = \begin{pmatrix}
																															\bm{U}_{11,t} & \bm{U}_{12,t} \\
																																	  \bm{0}  & \bm{U}_{22,t}
																													 \end{pmatrix} \begin{matrix} k \hfill \\ t-k \end{matrix},
\end{equation}
\label{eq:TVPGQRD1}%
\end{subequations}
where \( \bm{R}_t \in \Rset^{k \times k} \), \( \bm{U}_t \in \Rset^{t \times t} \) are upper triangular and non-singular and \( \bm{Q}_t, \bm{P}_{t,2} \) are orthogonal matrices of order \( t \). When \( \bm{Q}_t \) and \( \bm{P}_{t,2} \) are applied on \refeq{GLLSPt1}, it gives
\begin{equation*}
\underset{\bm{\beta}_t, \bm{v}_t}{\text{argmin}} \ \ \left\| \bm{P}_{t,2}^T \bm{v}_t \right\|^2 \ \text{subject to} \ \ \bm{Q}_t^T\bm{y}_t = \bm{Q}_t^T\bm{X}_t \bm{\beta}_t + \bm{Q}_t^T\bm{C}_t \bm{P}_{t,2} \bm{P}_{t,2}^T \bm{v}_t.
\end{equation*}
The GLLSP \refeq{GLLSPt1} now becomes
\begin{equation}
\underset{\bm{\beta}_t, \bm{v}_{t,A}, \bm{v}_{t,B}}{\text{argmin}} \ \left\| \begin{pmatrix} \bm{v}_{t,A} \\ \bm{v}_{t,B} \end{pmatrix} \right\|^2 \ \text{subject to} \ 
\begin{pmatrix}
		\bm{y}_{t,A} \\
		\bm{y}_{t,B}
\end{pmatrix} = \begin{pmatrix}
										\bm{R}_{t} \\
										\bm{0}
								\end{pmatrix} \bm{\beta}_t + \begin{pmatrix}
																								\bm{U}_{11,t} & \bm{U}_{12,t} \\
																									    \bm{0}  & \bm{U}_{22,t}
																						 \end{pmatrix} \begin{pmatrix} 
																														\bm{v}_{t,A} \\ 
																														\bm{v}_{t,B} 
																												\end{pmatrix},
\label{eq:GLLSPt2}
\end{equation}
where the second part of the restrictions in \refeq{GLLSPt2} yields \( \bm{v}_{t,B} = \bm{U}_{22,t}^{-1} \bm{y}_{t,B} \). The GLLSP in \refeq{GLLSPt2} is then reduced to
\begin{equation*}
\underset{\bm{\beta}_t, \bm{v}_{t,A}}{\text{argmin}} \  \left\| \bm{v}_{t,A} \right\|^2 \ \text{subject to} \ 
		\bm{\tilde{y}}_{t,A}  = \bm{R}_{t} \bm{\beta}_t + \bm{U}_{11,t} 	\bm{v}_{t,A},
\label{eq:GLLSPt3}
\end{equation*}
where  \( \bm{\tilde{y}}_{t,A} = \bm{y}_{t,A} - \bm{U}_{12,t} \bm{v}_{t,B} \). The estimator for \( \bm{\beta}_t \) is derived by setting \( \bm{v}_{t,A} = \bm{0}\), in order to minimise the argument, and from the solution of the upper triangular system \( \bm{R}_t \bm{\beta}_t = \bm{\tilde{y}}_{t,A} \).

\section{Multivariate Time-Varying Parameter Model}\label{sec:RecursiveTVPSUR}

\indent\par
A more general case of the TVP model in \refeq{TVPmodelt} is a system of $G$ such regressions which are contemporaneously correlated\index{contemporaneous correlation}. That is, consider the time-varying parameter seemingly unrelated regressions (TVP-SUR) model
\begin{equation*}
\begin{pmatrix}
\psi_{1,t} \\
\vdots  \\
\psi_{G,t}
\end{pmatrix} = \begin{pmatrix}
										\bm{x}_{1,t} &         &       \\
												   & \ddots  &       \\
                           &         & \bm{x}_{G,t}
								\end{pmatrix} \begin{pmatrix}
																\bm{\beta}_{1,t} \\
																\vdots \\
																\bm{\beta}_{G,t}
															\end{pmatrix} + \begin{pmatrix}
																									\epsilon_{1,t} \\
																									\vdots \\
																									\epsilon_{G,t}
																						\end{pmatrix}, \ \ t=1,\dots,M 
\end{equation*}
and
\begin{equation*}
\begin{pmatrix}
	\bm{\beta}_{1,t} \\
	\vdots \\
	\bm{\beta}_{G,t}
\end{pmatrix} = \begin{pmatrix}
									\bm{\beta}_{1,t-1} \\
									\vdots \\
									\bm{\beta}_{G,t-1}
								\end{pmatrix}	+ \begin{pmatrix}
																	\bm{\eta}_{1,t} \\
																	\vdots \\
																	\bm{\eta}_{G,t}
																\end{pmatrix},																					
\end{equation*}
where $\bm{x}_{jt}\in\Rset^{k_j}$ is a row vector of explanatory variables for regression $j$ at time $t$, \( ( \epsilon_{1,t} \ \ \hdots \ \ \epsilon_{G,t} )^T \)  is a $G\times 1$ disturbance vector with zero mean and variance covariance matrix \( \bm{\Sigma} = \left[ \sigma_{ij} \right]_{i,j} \), $i,j = 1,\dots, G$. Moreover, $\bm{\eta}_{jt} \sim (\bm{0}, \sigma_{jj}\bm{\Sigma}_{j})$ and $\mathbb{E} ( \bm{\eta}_{jt} \bm{\eta}^T_{it}) = \bm{0} $ for $i\neq j$. As in the constant coefficients seemingly unrelated regressions (SUR) model, when $\sigma_{ij} \neq 0$ for $i \neq j$, efficiency will be gained if the estimation of the unknown parameters is executed in a system of the $G$ regressions \citep{zellner:62,davidson:04}. Furthermore, let \( K = \sum_{i=1}^G {k_i} \). 

Consider the $i$th regression of the system with all the available observations up to time $t$, that is, 
\begin{equation*}
\begin{pmatrix}
\psi_{i,1} \\
\vdots  \\
\psi_{i,t-1} \\
\psi_{i,t}
\end{pmatrix} = \begin{pmatrix}
										\bm{x}_{i,1}    \\
										\vdots     \\
										\bm{x}_{i,t-1} \\
                    \bm{x}_{i,t}
								\end{pmatrix} \bm{\beta}_{i,t} + \begin{pmatrix}
																										\epsilon_{i,1} \\
																										\vdots \\
																										\epsilon_{i,t-1} \\
																										\epsilon_{i,t}
																							\end{pmatrix} - \begin{pmatrix}
																																	\bm{x}_{i,1} & \cdots & \bm{x}_{i,1} \\
																																	\vdots  		 & \ddots & \vdots \\
																																	\bm{0}			 & \hdots & \bm{x}_{i,t-1} \\
																																	\bm{0}			 & \cdots & \bm{0}  
																												\end{pmatrix} \begin{pmatrix}
																																					\bm{\eta}_{i,2} \\
																																					\vdots \\
																																					\bm{\eta}_{i,t}
																																			\end{pmatrix}	
\end{equation*}
or in compact form as
\begin{equation*}
\bm{y}_{i,t} = \bm{X}_{i,t} \bm{\beta}_{it} + \bm{e}^*_{i,t}, \quad \bm{e}^*_{i,t} \sim \left( \bm{0}, \sigma_{ii} \left( \bm{I}_t + \bm{A}_{i,t} ( \bm{I}_{t-1} \otimes \bm{\Sigma}_i ) \bm{A}^T_{i,t} \right) \right),
\end{equation*}
where \(\bm{e}^*_{i,t} = \bm{e}_{i,t} - \bm{A}_{i,t} \bm{u}_{i,t}\) is defined as in \refeq{TVPmodelt}.
The TVP-SUR model is then given in matrix form, at time \( t \), by
\begin{equation*}
\begin{pmatrix}
\bm{y}_{1,t} \\
\bm{y}_{2,t} \\
\vdots \\
\bm{y}_{G,t}
\end{pmatrix} = \begin{pmatrix}
										\bm{X}_{1,t} &              &         &       \\
												         & \bm{X}_{2,t} &         &       \\
												         &              & \ddots  &       \\
                                 &              &         & \bm{X}_{G,t}
								\end{pmatrix} \begin{pmatrix}
																 \bm{\beta}_{1,t} \\
																 \bm{\beta}_{2,t} \\
																 \vdots \\
																 \bm{\beta}_{G,t}
															\end{pmatrix} + \begin{pmatrix}
																									\bm{e}^*_{1,t} \\
																									\bm{e}^*_{2,t} \\
																									\vdots \\
																									\bm{e}^*_{G,t}
																							\end{pmatrix}, 
\end{equation*}
or equivalently by
\begin{equation}
\text{vec}\left( \{ \bm{y}_{i,t} \} \right)=\left( \oplus_{i=1}^G \bm{X}_{i,t} \right) \text{vec}\left( \{ \bm{\beta}_{i,t} \} \right)  + \text{vec}( \{ \bm{e}^*_{i,t} \}),
\label{eq:TVPSUR1}
\end{equation}
where \( \bm{y}_{i,t} \in \Rset^t \) are the response vectors at time $t$, \( \bm{X}_{i,t} \in\Rset^{t \times k_i}\) are the exogenous matrices at time $t$ with full column rank, \( \bm{\beta}_{i,t} \in\Rset^{k_i}\) are the time-varying coefficients at time $t$ and \( \bm{e}^*_{i,t} \in\Rset^{t} \) are the disturbance terms, \( i = 1,\dots,G\). Note that \( \{ \cdot \} \) denotes a set of vectors and \( \oplus_{i=1}^G  \) is the direct sum which for notational convenience will be abbreviated by \( \oplus_i \). 
The error term in \refeq{TVPSUR1} has zero mean and variance covariance matrix
\begin{align}
\begin{split}
\bm{\Omega}_t &= \begin{pmatrix}
										\sigma_{11}\bm{\Omega}_{1,t} & \sigma_{12}\bm{I}_t          & \hdots & \sigma_{1G}\bm{I}_t \\
							               \sigma_{21}\bm{I}_t & \sigma_{22}\bm{\Omega}_{2,t} & \hdots & \sigma_{2G}\bm{I}_t \\
											                    \vdots &              \vdots          & \ddots & \vdots  \\
													   \sigma_{G1}\bm{I}_t & \sigma_{G2}\bm{I}_t          & \hdots & \sigma_{GG}\bm{\Omega}_{G,t}
								\end{pmatrix}, \\
					 &= \oplus_i \sigma_{ii} \bm{A}_{i,t} (\bm{I}_{t-1} \otimes \bm{\Sigma}_i ) \bm{A}^T_{i,t} + \bm{\Sigma} \otimes \bm{I}_t \\
					 &= \oplus_i \bm{C}_{i,t} \bm{C}_{i,t}^T + \bm{C} \bm{C}^T \otimes \bm{I}_t \\
					 &= \begin{pmatrix} \oplus_i \bm{C}_{i,t} & \bm{C} \otimes \bm{I}_t \end{pmatrix} \begin{pmatrix} \oplus_i \bm{C}_{i,t} & \bm{C} \otimes \bm{I}_t \end{pmatrix}^T
\end{split}
\label{eq:TVPSURvarcov}					
\end{align}
where \( \bm{\Omega}_{i,t} = \left( \bm{I}_t + \bm{A}_{i,t} (\bm{I}_{t-1} \otimes \bm{\Sigma}_i ) \bm{A}^T_{i,t} \right)\), \(\bm{C}_{i,t} = \sqrt{\sigma_{ii}} \bm{A}_{i,t} (\bm{I}_{t-1} \otimes \bm{C}_i ) \) and \( \bm{\Sigma}_i = \bm{C}_i \bm{C}_i ^T \) is the Cholesky decomposition of $\bm{\Sigma}_i$. The best linear unbiased estimator of the TVP-SUR model \refeq{TVPSUR1} is obtained from the solution of the GLLSP 
\begin{equation}
\begin{array}{l}
\underset{\bm{\beta}_{i,t}, \bm{v}_{i,t}}{\text{argmin}} \ \ \left\| \text{vec} ( \{ \bm{v}_{i,t} \} ) \right\|^2 \ \text{subject to} \\
\text{vec} ( \{ \bm{y}_{i,t} \} ) = (\oplus_i \bm{X}_{i,t}) \text{vec} ( \{ \bm{\beta}_{i,t} \} ) + ( \oplus_i \bm{C}_{i,t} \ \bm{C} \otimes \bm{I}_t ) \text{vec} ( \{ \bm{v}_{i,t} \} ),
\end{array}
\label{eq:SURGLLSP1}
\end{equation}
where \( \text{vec} ( \{ \bm{v}_{i,t} \} ) \sim ( 0 , \bm{I}_{ (t-1)K +Gt } ) \) is such that \( \text{vec}( \{ \bm{e}^*_{i,t} \}) =  ( \oplus_i \bm{C}_{i,t} \ \ \bm{C} \otimes \bm{I}_t ) \text{vec} ( \{ \bm{v}_{i,t} \} ) \). The solution of \refeq{SURGLLSP1} is derived by computing the RQD
\begin{equation}
\begin{pmatrix} 
\oplus_i \bm{C}_{i,t} & \bm{C} \otimes \bm{I}_t 
\end{pmatrix} \bm{\tilde{P}}_{t,1} = \begin{pmatrix}
															\bm{0} & \bm{\tilde{C}}_t
													\end{pmatrix},
\label{eq:SURRQD1}
\end{equation}
and the GQRD
\begin{subequations}
\begin{equation}
 \bm{\tilde{Q}}^T_{t} \begin{pmatrix}
					\oplus_i \bm{X}_{i,t} &  \text{vec} ( \{ \bm{y}_{i,t} \} )
			 \end{pmatrix} = \begin{pmatrix}
													\oplus_i \bm{R}_{i,t} & \text{vec} ( \{ \bm{y}_{i,tA} \} ) \\
													\bm{0}								& \text{vec} ( \{ \bm{y}_{i,tB} \} )
											 \end{pmatrix},
\label{eq:SURGQRD1a}%
\end{equation}
\begin{equation}
\left( \bm{\tilde{Q}}^T_{t} \bm{C}_t \right) \bm{\tilde{P}}_{t,2} = \bm{L}_{t} = \begin{pmatrix} 
																														\bm{L}_{11,t} & \bm{L}_{12,t} \\
																																 \bm{0}   & \bm{L}_{22,t}
																											 \end{pmatrix} \begin{matrix} K \hfill \\ Gt -K, \end{matrix}
\label{eq:SURGQRD1b}%
\end{equation}
\label{eq:SURGQRD1}%
\end{subequations}
where \( \bm{\tilde{C}}_t \in \Rset^{ Gt \times Gt } \), \( \bm{R}_{i,t} \in \Rset^{k_i \times k_i} \), \( i = 1,\cdots,G\), and \( \bm{L}_{t} \in \Rset^{ Gt \times Gt } \) are upper triangular and non-singular, and \(  \bm{\tilde{P}}_{t,1} \in \Rset^{ Gt \times ( (t-1)K + Gt )} \), \( \bm{\tilde{Q}}^T_{t}, \bm{\tilde{P}}_{t,2}, \in \Rset^{ Gt \times Gt } \) are orthogonal. Note that in the above computations the special structure of the matrices is exploited. Numerically stable and computationally efficient strategies which exploit the special sparse structure of the matrices have been previously developed \citep{kontog:95,foschi:03,yanev:07}. Using the computations in \refeq{SURRQD1} and \refeq{SURGQRD1}, the GLLSP in \refeq{SURGLLSP1} is equivalently given by
\begin{equation}
\begin{array}{l}
\underset{\bm{\beta}_{i,t}, \bm{v}_{i,tA}, \bm{v}_{i,tB} }{\text{argmin}} \ \ \left\| \begin{pmatrix} \text{vec} ( \{ \bm{v}_{i,tA} \} ) \\ \text{vec} ( \{ \bm{v}_{i,tB} \} )  \end{pmatrix} \right\|^2 \ \text{subject to} \\
\\
\begin{pmatrix}
\text{vec} ( \{ \bm{y}_{i,tA} \} ) \\
\text{vec} ( \{ \bm{y}_{i,tB} \} ) 
\end{pmatrix} = \begin{pmatrix}
									\oplus_i \bm{R}_{i,t} \\
									\bm{0}
								\end{pmatrix} \text{vec} ( \{ \bm{\beta}_{i,t} \} ) + 
								\begin{pmatrix} 
									\bm{L}_{11,t} & \bm{L}_{12,t} \\
									   \bm{0}   & \bm{L}_{22,t}
								\end{pmatrix} \begin{pmatrix}
																	\text{vec} ( \{ \bm{v}_{i,tA} \} ) \\
																	\text{vec} ( \{ \bm{v}_{i,tB} \} ) 																
															\end{pmatrix},
\label{eq:TVPSURGLLSP2}
\end{array}
\end{equation}
where 
\begin{equation*}
\begin{pmatrix}
		\text{vec} ( \{ \bm{v}_{i,tA} \} ) \\
		\text{vec} ( \{ \bm{v}_{i,tB} \} ) \\																	
\end{pmatrix} = \bm{\tilde{P}}_{t,2}^T \bm{\tilde{P}}_{t,1} ^T \text{vec} ( \{ \bm{v}_{i,t} \} ).
\end{equation*}
The solution of the GLLSP in \refeq{TVPSURGLLSP2} is obtained by solving the triangular system 
\(
 \bm{L}_{22,t} \text{vec} ( \{ \bm{v}_{i,tB} \} ) = \text{vec} ( \{ \bm{y}_{i,tB} \} )
\)
for \( \bm{v}_{i,tB} \) and by setting \( \text{vec} ( \{ \bm{v}_{i,tA} \} ) = \bm{0} \) in order to minimise the argument in \refeq{TVPSURGLLSP2}. The BLUE of \( \bm{\beta}_{i,t} \), \( i = 1,\dots,G\), is derived from the solution of the triangular system
\(\oplus_i \bm{R}_{i,t} \text{vec} ( \{ \bm{\beta}_{i,t} \} ) = \text{vec} ( \{ \bm{\tilde{y}}_{i,tA} \} ) \),
where \( \text{vec} ( \{ \bm{\tilde{y}}_{i,tA} \} ) = \text{vec} ( \{ \bm{y}_{i,tA} \} ) - \bm{L}_{12,t} \text{vec} ( \{ \bm{v}_{i,tB} \} )\).
The steps for the numerical strategy for estimating the TVP-SUR model using orthogonal transformations are summarised in Algorithm~\refalg{AlgTVPSUR}.

\begin{algorithm}[H]
\caption{Estimating the TVP-SUR model (\ref{eq:TVPSUR1}) using orthogonal transformations.}
\label{Ta:AlgTVPSUR}
\begin{algorithmic}[1]
\algsetup{linenodelimiter=.}
\STATE Let \( \bm{\tilde{y}}_{i,t}, \bm{X}_{i,t}, \bm{C}_{i,t}\) and \( \bm{C} \). 
\STATE Compute the RQD \( \left(  \oplus_i \bm{C}_{i,t}  \ \ \bm{C} \otimes \bm{I}_t \right) \bm{P}_{t,1} = ( \bm{0} \ \ \bm{\tilde{C}}_{t}) \).
\STATE Compute the QRD in \refeq{SURGQRD1a}. 
\STATE Compute the RQD in \refeq{SURGQRD1b}. 
\STATE Compute \( \text{vec} ( \{ \bm{\tilde{y}}_{i,tA} \} ) = \text{vec} ( \{ \bm{y}_{i,tA} \} ) - \bm{L}_{12,t} \text{vec} ( \{ \bm{v}_{i,tB} \} )\). 
\STATE Solve the triangular system \(\oplus_i \bm{R}_{i,t} \text{vec} ( \{ \bm{\beta}_{i,t} \} ) = \text{vec} ( \{ \bm{\tilde{y}}_{i,tA} \} ) \) for \( \bm{\beta}_{i,t} \).
\end{algorithmic}
\end{algorithm} 

\subsection{Updating the TVP-SUR Model with one new Observation}\label{sec:UpdateTVPSURmodel}

\indent\par
Consider now updating each regression in the TVP-SUR model when a new datum is collected. This is defined as the original model \refeq{TVPSUR1} together with a single new observation in each regression which at time $t+1$ is given by
\begin{equation*}
\begin{array}{l}
\begin{pmatrix}
\psi_{1,t+1} \\
\vdots  \\
\psi_{G,t+1}
\end{pmatrix} = \begin{pmatrix}
										\bm{x}_{1,t+1} &         &       \\
												   & \ddots  &       \\
                           &         & \bm{x}_{G,t+1}
								\end{pmatrix} \begin{pmatrix}
																\bm{\beta}_{1,t+1} \\
																\vdots \\
																\bm{\beta}_{G,t+1}
															\end{pmatrix} + \begin{pmatrix}
																									\epsilon_{1,t+1} \\
																									\vdots \\
																									\epsilon_{G,t+1}
																							\end{pmatrix},  \\
																							\\
\begin{pmatrix}
	\bm{\beta}_{1,t+1} \\
	\vdots \\
	\bm{\beta}_{G,t+1}
\end{pmatrix} = \begin{pmatrix}
									\bm{\beta}_{1,t} \\
									\vdots \\
									\bm{\beta}_{G,t}
								\end{pmatrix}	+ \begin{pmatrix}
																	\bm{\eta}_{1,t+1} \\
																	\vdots \\
																	\bm{\eta}_{G,t+1}
																\end{pmatrix}.																					
\end{array}
\end{equation*}
The updated TVP-SUR model at time $t+1$ is written as
\begin{equation}
\begin{pmatrix}
\bm{y}_{1,t} \\
\psi_{1,t+1} \\
\bm{y}_{2,t} \\
\psi_{2,t+1} \\
\vdots \\
\bm{y}_{G,t} \\
\psi_{G,t+1}
\end{pmatrix} = \begin{pmatrix}
										\bm{X}_{1,t} &              &         &       \\
										\bm{x}_{1,t+1} &              &         &       \\
												         & \bm{X}_{2,t} &         &       \\
												         & \bm{x}_{2,t+1} &         &       \\
												         &              & \ddots  &       \\
                                 &              &         & \bm{X}_{G,t} \\
                                 &              &         & \bm{x}_{G,t+1}
								\end{pmatrix} \begin{pmatrix}
																 \bm{\beta}_{1,t+1} \\
																 \bm{\beta}_{2,t+1} \\
																 \vdots \\
																 \bm{\beta}_{G,t+1}
															\end{pmatrix} + \begin{pmatrix}
																									\bm{e}^*_{1,t} \\
																									\epsilon^*_{1,t+1} \\
																									\bm{e}^*_{2,t} \\
																									\epsilon^*_{2,t+1} \\
																									\vdots \\
																									\bm{e}^*_{G,t} \\
																									\epsilon^*_{G,t+1}
																							\end{pmatrix}, 
\label{eq:MTVPadd1}
\end{equation}
where the variance covariance matrix is 
\begin{equation*}
\bm{\Omega}_{t+1} = \begin{pmatrix}
							\sigma _{11} \bm{\Omega}_{1,t+1} & \sigma_{12} \bm{I}_{t+1}              & \hdots & \sigma_{1G} \bm{I}_{t+1}  \\
													 \sigma_{21} \bm{I}_{t+1}  & \sigma_{22} \bm{\Omega}_{2,t+1} & \hdots & \sigma_{2G} \bm{I}_{t+1}  \\
																						  \vdots & \vdots                                & \ddots & \vdots  \\
										       \sigma_{G1} \bm{I}_{t+1}  & \sigma_{G2} \bm{I}_{t+1}              & \hdots & \sigma_{GG}\bm{\Omega}_{G,t+1}
					 \end{pmatrix}
\end{equation*}
and \( \bm{\Omega}_{i,t+1} = \left( \bm{I}_{t+1} + \bm{A}_{i,t+1} (\bm{I}_{t} \otimes \bm{\Sigma}_i ) \bm{A}^T_{i,t+1} \right)\). Notice that the dispersion matrix of each time-varying regression is also updated by \( \bm{X}_{i,t} \bm{\Sigma}_i \bm{X}_{i,t}^T \) to encapsulate the new information available, namely,
\begin{equation*} 
\bm{\Omega}_{i,t+1} = \begin{pmatrix}
																\bm{\tilde{\Omega}}_{i,t} & \bm{0} \\
															                     \bm{0} & 1
														  \end{pmatrix} 
														= \begin{pmatrix}
																\bm{\Omega}_{i,t} + \bm{X}_{i,t} \bm{\Sigma}_i \bm{X}_{i,t}^T & \bm{0} \\
															                                                         \bm{0} & 1
														  \end{pmatrix}.
\label{eq:Omega.i.tilde}
\end{equation*}

For the recursive estimation of the TVP-SUR model, consider re-arranging the observations of the updated TVP-SUR model \refeq{MTVPadd1} as follows
\begin{equation}
\begin{pmatrix}
\bm{y}_{1,t} \\
\bm{y}_{2,t} \\
\vdots \\
\bm{y}_{G,t} \\
\psi_{1,t+1} \\
\psi_{2,t+1} \\
\vdots \\
\psi_{G,t+1}
\end{pmatrix} = \begin{pmatrix}
										\bm{X}_{1,t} &              &         &       \\
												         & \bm{X}_{2,t} &         &       \\
												         &              & \ddots  &       \\
                                 &              &         & \bm{X}_{G,t} \\
									\bm{x}_{1,t+1} &                &         &       \\
												         & \bm{x}_{2,t+1} &         &       \\
												         &                & \ddots  &       \\
                                 &                &         & \bm{x}_{G,t+1}
								\end{pmatrix} \begin{pmatrix}
																 \bm{\beta}_{1,t+1} \\
																 \bm{\beta}_{2,t+1} \\
																 \vdots \\
																 \bm{\beta}_{G,t+1}
															\end{pmatrix} + \begin{pmatrix}
																									\bm{\tilde{e}}^*_{1,t} \\
																									\bm{\tilde{e}}^*_{2,t} \\
																									\vdots \\
																									\bm{\tilde{e}}^*_{G,t} \\
																									\epsilon^*_{1,t+1} \\
																									\epsilon^*_{2,t+1} \\
																									\vdots \\
																									\epsilon^*_{G,t+1}
																							\end{pmatrix}, 
\label{eq:MTVPadd1permuted}
\end{equation}
which is conformably written as
\begin{equation*}
\begin{pmatrix}
\text{vec} ( \{ \bm{y}_{i,t} \} ) \\
\text{vec} ( \{ \psi_{i,t+1} \} )
\end{pmatrix} = \begin{pmatrix}
										\oplus_i \bm{X}_{i,t} \\
										\oplus_i \bm{x}_{i,t+1}
								\end{pmatrix} \text{vec} ( \{ \bm{\beta}_{i,t+1} \} ) + \begin{pmatrix}																													
																																						\text{vec} ( \{ \bm{\tilde{e}}_{i,t} \} ) \\
																																						\text{vec} ( \{ \epsilon^*_{i,t+1} \} )
																																				\end{pmatrix}, \quad
																																				\begin{pmatrix}																													
																																						\text{vec} ( \{ \bm{\tilde{e}}_{i,t} \} ) \\
																																						\text{vec} ( \{ \epsilon^*_{i,t+1} \} )
																																				\end{pmatrix} \sim \left( \bm{0} ,  \bm{\Omega}^*_{t+1} \right).
\end{equation*}
Now $ \bm{\Omega}^*_{t+1}$ is given by
\begin{equation*}
\bm{\Omega}^*_{t+1}  = \begin{pmatrix}
											\bm{\tilde{\Omega}}_{t} & \bm{0} \\
											\bm{0} & \bm{\Sigma}
									 \end{pmatrix},
\end{equation*}
where \( \bm{\tilde{\Omega}}_{t} \) is the updated variance covariance matrix of the first $t$ observations. That is, \refeq{TVPSURvarcov} is now revised to become 
\begin{align*}
\begin{split}
\bm{\tilde{\Omega}}_{t} & = \begin{pmatrix}
																\sigma _{11} \bm{\tilde{\Omega}}_{1,t}  & \sigma_{12} \bm{I}_{t}  & \hdots & \sigma_{1G} \bm{I}_{t}   \\
																\sigma_{21} \bm{I}_{t} & \sigma_{22} \bm{\tilde{\Omega}}_{2,t}  & \hdots & \sigma_{2G} \bm{I}_{t}  \\
																\vdots & \vdots           & \ddots & \vdots   \\
																\sigma_{G1} \bm{I}_{t} & \sigma_{G2} \bm{I}_{t} & \hdots & \sigma_{GG}\bm{\tilde{\Omega}}_{G,t}  	
														\end{pmatrix} \\
& = \bm{\Omega}_t + \oplus_i \bm{X}_{i,t} \bm{\Sigma}_i \bm{X}_{i,t}^T \\
& = \bm{\tilde{C}}_t \bm{\tilde{C}}_t^T + \oplus_i \bm{X}_{i,t} \bm{C}_{i} \bm{C}_{i}^T \bm{X}_{i,t}^T \\
& = \begin{pmatrix} 
			\bm{\tilde{C}}_t & \oplus_i \bm{X}_{i,t} \bm{C}_{i} 
		\end{pmatrix} \begin{pmatrix} 
											\bm{\tilde{C}}_t & \oplus_i \bm{X}_{i,t} \bm{C}_{i} 
									\end{pmatrix}^T,
\end{split}
\end{align*}
where \( \bm{\tilde{C}}_t \) is from the RQD in \refeq{SURRQD1} and $\bm{C}_i$ is the Cholesky factor of $\bm{\Sigma}_i$.
Then it follows that 
\begin{equation*}
\bm{\Omega}^*_{t+1} 
							  = \begin{pmatrix} 
											\bm{\tilde{C}}_t & \oplus_i \bm{X}_{i,t} \bm{C}_{i} & \bm{0} \\
											 \bm{0}  &      \bm{0}                    & \bm{C}
									 \end{pmatrix} 
									 \begin{pmatrix} 
											\bm{\tilde{C}}_t & \oplus_i \bm{X}_{i,t} \bm{C}_{i} & \bm{0} \\
											\bm{0}   & \bm{0}                         & \bm{C}
									 \end{pmatrix}^T.
\end{equation*}
Hence the GLLSP, which yields the BLUE of the updated by one observation TVP-SUR model, is given by
\begin{equation}
\begin{array}{l}
\underset{\tilde{\beta}_{i,t+1}, \bm{v}_{i,t}, \bm{v}^*_{i,t},\bm{v}_{t+1}}{\text{argmin}} \  \left\| \begin{pmatrix} \text{vec} ( \{ \bm{v}_{i,t} \} ) \\ 	\text{vec} ( \{ \bm{v}^*_{i,t} \} ) \\ 	\bm{v}_{t+1}  \end{pmatrix} \right\|^2 \ \text{subject to} \\
\\
\begin{pmatrix}
\text{vec} ( \{ \bm{y}_{i,t} \} ) \\
\text{vec} ( \{ \psi_{i,t+1} \} )
\end{pmatrix} = \begin{pmatrix}
										\oplus_i \bm{X}_{i,t} \\
										\oplus_i \bm{x}_{i,t+1}
								\end{pmatrix} \text{vec} ( \{ \bm{\beta}_{i,t+1} \} ) + 
								\begin{pmatrix} 
			\bm{\tilde{C}}_t & \oplus_i \bm{X}_{i,t} \bm{C}_{i} & \bm{0} \\
			 \bm{0}  &      \bm{0}                    & \bm{C}
		\end{pmatrix}
								\begin{pmatrix}																													
										\text{vec} ( \{ \bm{v}_{i,t} \} ) \\
										\text{vec} ( \{ \bm{v}^*_{i,t} \} ) \\
										\bm{v}_{t+1}
								\end{pmatrix},
\end{array}
\label{eq:UTVPSURGLLSP}
\end{equation}
where previous computations from the solution of the GLLSP \refeq{SURGLLSP1} can be efficiently utilised to reduce the computational cost. Namely, using the GQRD in \refeq{SURRQD1} and \refeq{SURGQRD1} and the solution of \refeq{TVPSURGLLSP2}, the latter GLLSP becomes
\begin{equation*}
\begin{array}{l}
\underset{\tilde{\beta}_{i,t+1}, \bm{v}_{i,tA}, \bm{v}_{i,tB}, \bm{v}^*_{i,t},\bm{v}_{t+1}}{\text{argmin}} \  \left\| \begin{pmatrix} \text{vec} ( \{ \bm{v}_{i,tA} \} ) \\ \text{vec} ( \{ \bm{v}_{i,tB} \} )  \\	\text{vec} ( \{ \bm{v}^*_{i,t} \} ) \\ 	\bm{v}_{t+1}  \end{pmatrix} \right\|^2 \ \text{subject to} \\
\\
\begin{pmatrix}
\text{vec} ( \{ \bm{y}_{i,tA} \} ) \\
\text{vec} ( \{ \bm{y}_{i,tB} \} ) \\
\text{vec} ( \{ \psi_{i,t+1} \} )
\end{pmatrix} = \begin{pmatrix}
										\oplus_i \bm{R}_{i,t} \\
										\bm{0} \\
										\oplus_i \bm{x}_{i,t+1}
								\end{pmatrix} \text{vec} ( \{ \bm{\beta}_{i,t+1} \} ) + 
								\begin{pmatrix} 
										\bm{L}_{11,t} & \bm{L}_{12,t} & \oplus_i \bm{R}_{i,t} \bm{C}_{i} & \bm{0} \\
											   \bm{0} & \bm{L}_{22,t} & \bm{0}                         & \bm{0} \\
											   \bm{0} & \bm{0}      & \bm{0}                         & \bm{C}	
		\end{pmatrix}
								\begin{pmatrix}																													
										\text{vec} ( \{ \bm{v}_{i,tA} \} ) \\
										\text{vec} ( \{ \bm{v}_{i,tB} \} ) \\
										\text{vec} ( \{ \bm{v}^*_{i,t} \} ) \\
										\bm{v}_{t+1}
								\end{pmatrix},
\end{array}
\end{equation*}
which reduces to
\begin{equation}
\begin{array}{l}
\underset{\bm{\tilde{\beta}}, \bm{v}_{i,tA}, \bm{v}^*_{i,t}, \bm{v}_{t+1}}{\text{argmin}} \  \left\| 
\begin{pmatrix} \text{vec} ( \{ \bm{v}_{i,tA} \} ) \\ \text{vec} ( \{ \bm{v}^*_{i,t} \} ) \\ \bm{v}_{t+1}  \end{pmatrix} \right\|^2 \ \text{subject to} \\
\\
\begin{pmatrix}
\text{vec} ( \{ \bm{\tilde{y}}_{i,tA} \} ) \\
\text{vec} ( \{ \psi_{i,t+1} \} )
\end{pmatrix} = \begin{pmatrix}
										\oplus_i \bm{R}_{i,t} \\
										\oplus_i \bm{x}_{i,t+1}
								\end{pmatrix} \text{vec} ( \{ \bm{\beta}_{i,t+1} \} ) + 
								\begin{pmatrix} 
										\bm{L}_{11,t} & \oplus_i \bm{R}_{i,t} \bm{C}_{i} & \bm{0} \\
											   \bm{0} & \bm{0}                         & \bm{C}												
		\end{pmatrix}
								\begin{pmatrix}																													
										\text{vec} ( \{ \bm{v}_{i,tA} \} ) \\
										\text{vec} ( \{ \bm{v}^*_{i,t} \} ) \\
										\bm{v}_{t+1}
								\end{pmatrix},
\end{array}
\label{eq:UTVPSURGLLSP1}
\end{equation}
where \( \bm{\tilde{y}}_{i,tA} = \bm{y}_{i,tA} - \bm{L}_{12,t} \bm{v}_{i,tB} \). The GLLSP in \refeq{UTVPSURGLLSP1} is solved in two stages. Firstly, by computing the updating RQD 
\begin{equation}
\begin{pmatrix} 
		\bm{L}_{11,t} & \oplus_i \bm{R}_{i,t} \bm{C}_{i} 
\end{pmatrix} \bm{P}_{t+1,1} = \begin{pmatrix}
																	\bm{\tilde{L}}_{11,t} & \bm{0}
															 \end{pmatrix},
\label{eq:TVPSUR.URQD1}
\end{equation}
where \( \bm{\tilde{L}}_{11,t} \in \Re^{K \times K}\) is upper triangular and non-singular, and \( \bm{P}_{t+1,1} \in \Re^{ 2K \times 2K}\). Employing \refeq{TVPSUR.URQD1} in \refeq{UTVPSURGLLSP1} yields the equivalent GLLSP
\begin{equation}
\begin{array}{l}
\underset{\bm{\tilde{\beta}}, \bm{\tilde{v}}_{i,tA}, \bm{v}_{t+1}}{\text{argmin}} \  \left\| 
\begin{pmatrix} \text{vec} ( \{ \bm{\tilde{v}}_{i,tA} \} ) \\ \bm{v}_{t+1}  \end{pmatrix} \right\|^2 \ \text{subject to} \\
\\
\begin{pmatrix}
\text{vec} ( \{ \bm{\tilde{y}}_{i,tA} \} ) \\
\text{vec} ( \{ \psi_{i,t+1} \} )
\end{pmatrix} = \begin{pmatrix}
										\oplus_i \bm{R}_{i,t} \\
										\oplus_i \bm{x}_{i,t+1}
								\end{pmatrix} \text{vec} ( \{ \bm{\beta}_{i,t+1} \} ) + 
								\begin{pmatrix} 
										\bm{\tilde{L}}_{11,t} & \bm{0} \\
											     \bm{0} & \bm{C}												
		\end{pmatrix}
								\begin{pmatrix}																													
										\text{vec} ( \{ \bm{\tilde{v}}_{i,tA} \} ) \\
										\bm{v}_{t+1}
								\end{pmatrix}.
\end{array}
\label{eq:UTVPSURGLLSP2}
\end{equation}
Secondly, by computing the updating GQRD
\begin{subequations}
\begin{equation}
\bm{Q}^T_{t+1} \begin{pmatrix}
		\oplus_i \bm{R}_{i,t} & \text{vec} ( \{ \bm{\tilde{y}}_{i,tA} \} ) \\
		\oplus_i \bm{x}_{i,t+1} & \text{vec} ( \{ \psi_{i,t+1} \} )
\end{pmatrix} = \begin{pmatrix}
									\oplus_i \bm{R}_{i,t+1} & \text{vec} ( \{ \bm{y}_{i,t+1A} \} ) \\
									\bm{0}                  & \text{vec} ( \{ \psi_{i,t+1B} \} )
								\end{pmatrix},
\label{eq:TVPSUR.UGQRDa}
\end{equation}
\begin{equation}
\bm{Q}^T_{t+1} 
\begin{pmatrix} 
		\bm{\tilde{L}}_{11,t} & \bm{0} \\
									 \bm{0} & \bm{C}												
\end{pmatrix} \bm{P}_{t+1,2} = \bm{L}_{t+1} = 
															\begin{pmatrix} 
																	\bm{L}_{11,t+1} & \bm{L}_{12,t+1} \\
																				   \bm{0} & \bm{L}_{22,t+1}												
															 \end{pmatrix},
\label{eq:TVPSUR.UGQRDb}
\end{equation}
\label{eq:TVPSUR.UGQRD}%
\end{subequations}
where \( \bm{R}_{i,t} \in \Re^{k_i \times k_i} \), $i=1,\dots,G$, \( \bm{L}_{t+1} \in \Re^{K \times K} \) are upper triangular and non-singular and \( \bm{Q}_{t+1} \), \( \bm{P}_{t+1,2} \) are orthogonal matrices of order $K+G$. The GLLSP is now given by
\begin{equation*}
\begin{array}{l}
\underset{\bm{\tilde{\beta}}_{i,t+1}, \bm{v}_{i,t+1A}, \bm{v}_{t+1B}}{\text{argmin}} \  \left\| \begin{pmatrix} \text{vec} ( \{ \bm{v}_{i,t+1A} \} ) \\ 	\bm{v}_{i,t+1B}  \end{pmatrix} \right\|^2 \ \text{subject to} \\
\\
\begin{pmatrix}
\text{vec} ( \{ \bm{y}_{i,t+1A} \} ) \\
\text{vec} ( \{ \psi_{i,t+1B} \} )
\end{pmatrix} = \begin{pmatrix}
										\oplus_i \bm{R}_{i,t+1} \\
									  \bm{0}
								\end{pmatrix} \text{vec} ( \{ \bm{\beta}_{i,t+1} \} ) + 
								\begin{pmatrix} 
										\bm{L}_{11,t+1} & \bm{L}_{12,t+1} \\
														 \bm{0} & \bm{L}_{22,t+1}										
								\end{pmatrix}
								\begin{pmatrix}																													
										\text{vec} ( \{ \bm{v}_{i,t+1A} \} ) \\
										\bm{v}_{t+1B}
								\end{pmatrix},
\end{array}
\end{equation*}
where \( 	(	\text{vec} ( \{ \bm{v}_{i,t+1A} \} )^T \ \ \bm{v}_{t+1B}^T )^T = \bm{P}^T_{t+1,2} ( (\text{vec} ( \{ \bm{v}_{i,tA} \} )^T \ \ \text{vec} ( \{ \bm{v}^*_{i,t} \} )^T ) \bm{P}_{t+1,1} \ \ \bm{v}_{t+1}^T )^T \). The latter GLLSP is solved in a similar way to \refeq{TVPSURGLLSP2}. The strategy for updating the TVP-SUR model with a single new observation is summarised in Algorithm~\refalg{AlgUTVPSUR}.

\begin{algorithm}[H]
\caption{Estimating the updated TVP-SUR model (\ref{eq:MTVPadd1}) using orthogonal transformations.}
\label{Ta:AlgUTVPSUR}
\begin{algorithmic}[1]
\algsetup{linenodelimiter=.}
\STATE Let \( \bm{\tilde{y}}_{i,tA}, \bm{R}_{i,t}, \bm{L}_{11,t} \) and \( \bm{P}_{t,2} \) emanate from the solution of the GLLSP \refeq{SURGLLSP1}.
\STATE Compute the RQD \( \left( \bm{L}_{11,t} \ \  \oplus_i \bm{R}_{i,t} \bm{C}_{i} \right) \bm{P}_{t+1,1} = ( \bm{\tilde{L}}_{11,t} \ \ \bm{0} ) \).
\STATE Compute the updating QRD in \refeq{TVPSUR.UGQRDa}. 
\STATE Compute the updating RQD in \refeq{TVPSUR.UGQRDb}. 
\STATE Compute \( \text{vec} ( \{ \bm{\tilde{y}}_{i,t+1A} \} )  = \text{vec} ( \{ \bm{y}_{i,t+1A} \} ) - \bm{L}_{12,t+1} \bm{v}_{t+1B} \). 
\STATE Solve the triangular system \( \oplus_i \bm{R}_{i,t+1} \text{vec} ( \{ \bm{\beta}_{i,t+1} \} ) = \text{vec} ( \{ \bm{\tilde{y}}_{i,t+1A} \} ) \) for \( \bm{\beta}_{i,t+1} \).
\end{algorithmic}
\end{algorithm}

\subsection{Multivariate Smoothing}\label{sec:SmoothingTVPSURmodel}

\indent\par
Consider now estimating \( \bm{\beta}_{i,t} \) based on information up to time $M$, $M>t$. That is, at time $M$ the estimates of $\bm{\beta}_{i,t}$ will be re-estimated in order to be revised given the full sample of data. In a way similar to the filtering of the TVP model, it obtains that 
\begin{equation*}
\bm{\beta}_M = \bm{\beta}_{M-1} + \bm{\eta}_M = \dots= \bm{\beta}_t + \sum_{s=t+1}^M \bm{\eta}_s . 
\end{equation*}
In order to derive the smoothing estimate of \( \bm{\beta}_{i,t} \), say \( \bm{\beta}_{i,t|M} \), given the full sample, consider the following system of observations for each time-varying regression 
\begin{equation}
\begin{pmatrix}
\psi_{i,t+1} \\
\psi_{i,t+2} \\
\vdots \\
\psi_{i,M}
\end{pmatrix} = \begin{pmatrix}
										\bm{x}_{i,t+1} \\
										\bm{x}_{i,t+2} \\
										\vdots \\
										\bm{x}_{i,M}
								\end{pmatrix} \bm{\beta}_{t|M} + \begin{pmatrix}
																								\bm{\epsilon}_{t+1} \\
																								\bm{\epsilon}_{t+2} \\
																								\vdots \\
																								\bm{\epsilon}_M
																				\end{pmatrix} + \begin{pmatrix}
																														 \bm{x}_{t+1} & \bm{0}   		 & \cdots & \bm{0}  \\
																														 \bm{x}_{t+2} & \bm{x}_{t+2} & \cdots & \bm{0}  \\
																														 \vdots  			& \vdots   		 & \ddots & \vdots  \\
																														 \bm{x}_M     & \bm{x}_M     & \cdots & \bm{x}_M
																												\end{pmatrix} \begin{pmatrix}
																																					\bm{\eta}_{t+1} \\
																																					\bm{\eta}_{t+2} \\
																																					\vdots \\
																																					\bm{\eta}_M
																																			\end{pmatrix}.			
\label{eq:SmoothTVPSUR}
\end{equation}
The latter is given in compact form by
\begin{equation}
\bm{y}_{t+1:M} = \bm{X}_{t+1:M} \bm{\beta}_{t|M} + \bm{e}_{t+1:M} + \bm{\tilde{A}}_{t+1:M} \bm{u}_{t+1:M}, 
\label{eq:TVPmodeltp1M}
\end{equation}
where the error term \( \bm{e}_{t+1:M} + \bm{\tilde{A}}_{t+1:M} \bm{u}_{t+1:M} \) has zero mean and variance covariance matrix \( \sigma^2 \bm{\tilde{\Omega}}_{t+1:M} = \sigma^2 ( \bm{I}_{M-t} + \bm{\tilde{A}}_{t+1:M} (\bm{I}_{M-t} \otimes \bm{\Sigma}_{\eta}) \bm{\tilde{A}}^T_{t+1:M} ) \). 
The system of regressions in \refeq{SmoothTVPSUR} is used to form the following model 
\begin{equation}
\begin{array}{l}
\begin{pmatrix}
\text{vec} ( \{ \bm{y}_{i,t} \} ) \\
\text{vec} ( \{ \bm{y}_{i,t+1:M} \} )
\end{pmatrix} = \begin{pmatrix}
										\oplus_i \bm{X}_{i,t} \\
										\oplus_i \bm{X}_{i,t+1:M}
								\end{pmatrix} \text{vec} ( \{ \bm{\beta}_{i,t|M} \} ) + 
								\begin{pmatrix}																													
										\text{vec} ( \{ \bm{e}^*_{i,t} \} ) \\
										\text{vec} ( \{ \bm{e}^*_{i,t+1:M} \} )
								\end{pmatrix} \\ 				
								\\
											\text{with} \ \ 
								\begin{pmatrix}																													
										\text{vec} ( \{ \bm{e}^*_{i,t} \} ) \\
										\text{vec} ( \{ \bm{e}^*_{i,t+1:M} \} )
								\end{pmatrix} \sim \left( \bm{0} , 
								\begin{pmatrix}
											\bm{\Omega}_{t} & \bm{0} \\
											         \bm{0} & \bm{\Omega}_{t|t+1:M}
									 \end{pmatrix} \right),
\end{array}
\label{eq:SmoothingTVPSUR}
\end{equation}
where the first block of rows in \refeq{SmoothingTVPSUR} is the TVP-SUR model \refeq{TVPSUR1} used in obtaining the filtering estimates of the model up to time $t$. 
Also, the block diagonal elements are given by \( \bm{\Omega}_{i,t+1:M} =  \oplus_i \sigma_{ii} \bm{\tilde{A}}_{i,t+1:M} (\bm{I}_{M-t} \otimes \bm{\Sigma}_i ) \bm{\tilde{A}}^T_{i,t+1:M}\) and the off-diagonal elements are \( \sigma_{ij}\bm{I}_{M-t}, \ i,j=1,\dots,G \). Notice that \( \bm{\tilde{A}}_{i,t+1:M} \), \( i=1,\dots,G \), are analogous to \( \bm{\tilde{A}}_{t+1:M} \) in \refeq{TVPmodeltp1M}. Also let \( \bm{\Omega}_{t|t+1:M} = \bm{\tilde{C}}_{t+1:M} \bm{\tilde{C}}_{t+1:M}^T \) be the Cholesky factorisation of \( \bm{\Omega}_{t|t+1:M} \). 

The estimation problem of model \refeq{SmoothingTVPSUR} is now written as the GLLSP
\begin{equation}
\begin{array}{l}
\underset{\bm{\beta}_{i,t|M}, \bm{v}_{i,1:t}, \bm{v}_{i,t+1:M}}{\text{argmin}} \  \left\| \begin{pmatrix} \text{vec} ( \{ \bm{v}_{i,t} \} ) \\	\text{vec} ( \{ \bm{v}_{i,t+1:M} \} ) \end{pmatrix} \right\|^2 \ \text{subject to} \\
\\
\begin{pmatrix}
\text{vec} ( \{ \bm{y}_{i,t} \} ) \\
\text{vec} ( \{ \bm{y}_{i,t+1:M} \} )
\end{pmatrix} = \begin{pmatrix}
										\oplus_i \bm{X}_{i,t} \\
										\oplus_i \bm{X}_{i,t+1:M}
								\end{pmatrix} \text{vec} ( \{ \bm{\beta}_{i,t|M} \} ) + 
								\begin{pmatrix}
									  \bm{\tilde{C}}_t &  \bm{0} \\
										\bm{0}   &  \bm{\tilde{C}}_{t+1:M}
							  \end{pmatrix}	
								\begin{pmatrix}																													
										\bm{P}_{1,t}^T \text{vec} ( \{ \bm{v}_{i,t} \} ) \\
										\text{vec} ( \{ \bm{v}_{i,t+1:M} \} )
								\end{pmatrix}.
\end{array}
\label{eq:SmoothingTVPSURGLLSP1}
\end{equation}
Given \refeq{SURGQRD1} and \refeq{TVPSURGLLSP2}, it follows that the latter is equivalent to
\begin{equation}
\begin{array}{l}
\underset{\bm{\beta}_{i,t|M}, \bm{v}_{i,tA}, \bm{v}_{i,t+1:M}}{\text{argmin}} \  \left\| \begin{pmatrix} \text{vec} ( \{ \bm{v}_{i,tA} \} ) \\ \text{vec} ( \{ \bm{v}_{i,t+1:M} \} ) \end{pmatrix} \right\|^2 \ \text{subject to} \\
\\
\begin{pmatrix}
\text{vec} ( \{ \bm{\tilde{y}}_{i,tA} \} ) \\
\text{vec} ( \{ \bm{y}_{i,t+1:M} \} )
\end{pmatrix} = \begin{pmatrix}
										\oplus_i \bm{R}_{i,t} \\
										\oplus_i \bm{X}_{i,t+1:M}
								\end{pmatrix} \text{vec} ( \{ \bm{\beta}_{i,t|M} \} ) + 
								\begin{pmatrix}
									  \bm{L}_{11,t} &  \bm{0} \\
										     \bm{0}   &  \bm{\tilde{C}}_{t+1:M}
							  \end{pmatrix}	
								\begin{pmatrix}																													
										\text{vec} ( \{ \bm{v}_{i,tA} \} ) \\
										\text{vec} ( \{ \bm{v}_{i,t+1:M} \} )
								\end{pmatrix}.
\end{array}
\label{eq:GLLSPsmothing}
\end{equation}
The solution of \refeq{GLLSPsmothing} is analogous to that of the GLLSP in \refeq{TVPSURGLLSP2} and follows from the updating GQRD 
\begin{subequations}
\begin{equation}
\bm{\tilde{Q}}^T_{(s)} \begin{pmatrix}
										\oplus_i \bm{R}_{i,t}     & \text{vec} ( \{ \bm{\tilde{y}}_{i,tA} \} ) \\
										\oplus_i \bm{X}_{i,t+1:M} & \text{vec} ( \{ \bm{y}_{i,t+1:M} \} )
								\end{pmatrix}  = 
								\begin{pmatrix}
										\oplus_i \bm{R}_{i,s} & \text{vec} ( \{ \bm{\tilde{y}}_{i,(s)A} \} ) \\
										                    0 & \text{vec} ( \{ \bm{y}_{i,(s)B} \} )
								\end{pmatrix}
\label{eq:SmoothingTVPSUR.GQRDa}
\end{equation}
\text{and} 
\begin{equation}
\bm{\tilde{Q}}^T_{(s)} \begin{pmatrix}
									  \bm{L}_{11,t} &  \bm{0} \\
										     \bm{0}   &  \bm{\tilde{C}}_{t+1:M}
							  \end{pmatrix} \bm{\tilde{P}}_{(s)} = 
								\begin{pmatrix}
									  \bm{\tilde{L}}_{11,(s)} &  \bm{\tilde{L}}_{12,(s)} \\
										               \bm{0}   &  \bm{\tilde{L}}_{22,(s)}
							  \end{pmatrix}.
\label{eq:SmoothingTVPSUR.GQRDb}
\end{equation} 
\end{subequations}
Algorithm~\refalg{AlgSmoothingTVPSUR} below summarises the steps for obtaining the smoothing estimates using the proposed method.

\begin{algorithm}[H]
\caption{Computing the smoothing estimates of the TVP-SUR model (\ref{eq:SmoothingTVPSUR}).}
\label{Ta:AlgSmoothingTVPSUR}
\begin{algorithmic}[1]
\algsetup{linenodelimiter=.}
\STATE Let \( \bm{\tilde{y}}_{i,tA}, \bm{R}_{i,t}, \bm{L}_{11,t} \) and \( \bm{P}_{t,2} \) emanate from the solution of the GLLSP \refeq{SURGLLSP1}.
\STATE Compute the updating QRD \refeq{SmoothingTVPSUR.GQRDa}. 
\STATE Compute the RQD \refeq{SmoothingTVPSUR.GQRDb}. 
\STATE Let \( (	\text{vec} ( \{ \bm{\tilde{v}}_{s,A} \} )^T \ \ \bm{\tilde{v}}_{s,B}^T )^T = \bm{\tilde{P}}^T_{(s)} ( \text{vec} ( \{ \bm{v}_{i,tA} \} )^T \ \ \bm{v}_{t+1:M}^T )^T \).
\STATE Compute \( \text{vec} ( \{ \bm{\tilde{y}}_{i,(s)A} \} )  = \text{vec} ( \{ \bm{y}_{i,(s)A} \} ) - \bm{\tilde{L}}_{12,(s)} \bm{\tilde{v}}_{s,B} \). 
\STATE Solve the triangular system \( \oplus_i \bm{R}_{i,s} \text{vec} ( \{ \bm{\beta}_{i,t|M} \} ) = \text{vec} ( \{ \bm{\tilde{y}}_{i,(s)A} \} ) \) for \( \bm{\beta}_{i,t|M} \).
\end{algorithmic}
\end{algorithm}

\subsection{Window Estimation of the TVP-SUR Model}\label{sec:DownWindonTVPSUR}

\indent \par

While a model needs to be updated with the most recent data to keep the estimates up to date, often it is possible that observations will need to be removed from a model so that they no longer affect the estimation results. Observations are excluded from a model because they are old or because they have been detected to be outliers or influential data. Many a time, deleting observations from a model will occur in parallel with adding observations. This is part of the estimation over a rolling window of data and cross validation procedures. 

Assume that the TVP model \refeq{TVPSUR1} has been estimated and at time $t+1$ a rolling window moves forward acquiring one new observation and discarding the oldest one from the model.  
That is, consider \refeq{MTVPadd1} and partition \( \bm{y}_{i,t+1} \), \( \bm{X}_{i,t+1} \) and \( \bm{e}^*_{i,t+1} \) as follows 
\begin{equation*}
\bm{y}_{i,t+1} = \begin{pmatrix}
									\psi^{(d)}_{i} \\ 
									\bm{y}^{(r)}_{i} \\
									\psi^{(n)}_{i}
							 \end{pmatrix} \begin{matrix} 1 \hfill \\ t-1 \\ 1 \hfill \end{matrix}, \ \
\bm{X}_{i,t+1} = \begin{pmatrix}
									\bm{x}^{(d)}_{i} \\
									\bm{X}^{(r)}_{i} \\
									\bm{x}^{(n)}_{i}
							 \end{pmatrix} \ \ \text{and} \ \ 
\bm{e}^*_{i,t+1} = \begin{pmatrix}
									e^{(d)}_{i} \\
									\bm{e}^{(r)}_{i} \\
									e^{(n)}_{i}
							 \end{pmatrix},
\end{equation*}
where \( \bm{x}^{(d)}_{i} \) is the deleted observation from the $i$th regression of the  model, \( \bm{X}^{(r)}_{i} \) are the remaining observations in the model and \( \bm{x}^{(n)}_{i} \) is the new observation included in the model. Using the above partitioning, and applying a permutation of the model as in \refeq{MTVPadd1permuted}, 
\begin{equation}
\begin{array}{l}
\begin{pmatrix}
\text{vec} ( \{ \bm{y}^{(r)}_{i} \} ) \\
\text{vec} ( \{ \psi^{(n)}_{i} \} )
\end{pmatrix} = \begin{pmatrix}
										\oplus_i \bm{X}^{(r)}_{i} \\
										\oplus_i \bm{x}^{(n)}_{i}
								\end{pmatrix} \text{vec} ( \{ \bm{\bar{\beta}}_{i} \} ) + 
								\begin{pmatrix}
										\text{vec} ( \{ \bm{e}^{(r)}_{i} \} ) \\
										\text{vec} ( \{ e^{(n)}_{i} \} )
								\end{pmatrix}, \\
								\\
								\begin{pmatrix}
										\text{vec} ( \{ \bm{e}^{(r)}_{i} \} ) \\
										\text{vec} ( \{ e^{(n)}_{i} \} )
								\end{pmatrix} \sim \left( \bm{0} , \begin{pmatrix} 
																												 \bm{\bar{\Omega}}^{(r)} & \bm{0} \\
																												       \bm{0}    & \bm{\Sigma}
																									 \end{pmatrix} \right).
\end{array}
\label{eq:TVPSURwindow}
\end{equation}

For the sequential estimation of \( \bm{\bar{\beta}}_{i} \) over a window of data, consider the following TVP-SUR model
\begin{equation}
\begin{array}{l}
\begin{pmatrix}
\text{vec} ( \{ \imath \psi^{(d)}_{i} \} ) \\
\text{vec} ( \{ \bm{y}_{i,t} \} ) \\
\text{vec} ( \{ \psi^{(n)}_{i} \} )
\end{pmatrix} = \begin{pmatrix}
										\oplus_i \imath \bm{x}^{(d)}_{i} \\
										\oplus_i \bm{X}_{i,t} \\
										\oplus_i \bm{x}^{(n)}_{i}
								\end{pmatrix} \text{vec} ( \{ \bm{\bar{\beta}}_{i} \} ) + 
								\begin{pmatrix}
										\text{vec} ( \{ \imath e^{(d)}_{i} \} ) \\
										\text{vec} ( \{ \bm{e}_{i,t} \} ) \\
										\text{vec} ( \{ e^{(n)}_{i} \} )
								\end{pmatrix}, \\
								\\
								\begin{pmatrix}
										\text{vec} ( \{ \imath e^{(d)}_{i} \} ) \\
										\text{vec} ( \{ \bm{e}_{i,t} \} ) \\
										\text{vec} ( \{ e^{(n)}_{i} \} )
								\end{pmatrix} \sim \left( \bm{0} , \bar{\Omega}
																									\right).
\label{eq:TVPSURwindow2}
\end{array}
\end{equation}
The variance covariance matrix in \refeq{TVPSURwindow2} is given by
\begin{equation*}
\bar{\Omega} = \begin{pmatrix}
									\bm{C}_{d}  & \imath \bm{\tilde{C}}_{d,t-d}  & \oplus_i \imath \bm{x}^{(d)}_{i,t} \bm{C}_i  & \bm{0} \\
									     \bm{0} & \bm{\tilde{C}}_t               & \oplus_i \bm{X}_{i,t} \bm{C}_i               & \bm{0} \\
											 \bm{0} & \bm{0}                         & \bm{0}                                       & \bm{C}
							  \end{pmatrix}	\Phi 
								\begin{pmatrix}
									\bm{C}_{d}  & \imath \bm{\tilde{C}}_{d,t-d}  & \oplus_i \imath \bm{x}^{(d)}_{i,t} \bm{C}_i  & \bm{0} \\
									     \bm{0} & \bm{\tilde{C}}_t               & \oplus_i \bm{X}_{i,t} \bm{C}_i               & \bm{0} \\
											 \bm{0} & \bm{0}                         & \bm{0}                                       & \bm{C}
							  \end{pmatrix}	^T
\end{equation*}
and is such that the effect of the oldest observation is excluded from the current estimate but the new information from the acquired observation will be incorporated. The imaginary unit in \refeq{TVPSURwindow2} gives the weight needed to \textit{downdate} the model, that is, to eliminate the affect of the first datum \citep{hadjiantoni:16}. 
The window estimation problem is then given by
\begin{equation}
\begin{array}{l}
\underset{\tilde{\beta}, \bm{v}^{(d)}_{i,t}, \bm{v}_{i,t}, v^{(n)}_{i}}{\text{argmin}} \  \left\| \begin{pmatrix} \text{vec} ( \{ \imath \bm{v}^{(d)}_{i,t} \} ) \\ \text{vec} ( \{ \bm{v}_{i,t} \} ) \\ \text{vec} ( \{ v^{(n)}_{i} \} ) \end{pmatrix} \right\|_h \ \text{subject to} \\
\\
\begin{pmatrix}
\text{vec} ( \{ \imath \psi^{(d)}_{i} \} ) \\
\text{vec} ( \{ \bm{y}_{i,t} \} ) \\
\text{vec} ( \{ \psi^{(n)}_{i} \} )
\end{pmatrix} = \begin{pmatrix}
										\oplus_i \imath \bm{x}^{(d)}_{i} \\
										\oplus_i \bm{X}_{i,t} \\
										\oplus_i \bm{x}^{(n)}_{i}
								\end{pmatrix} \text{vec} ( \{ \bm{\bar{\beta}}_{i} \} ) + 
								\begin{pmatrix}
									\bm{C}_{d}  & \imath \bm{\tilde{C}}_{d,t-d}  & \oplus_i \imath \bm{x}^{(d)}_{i,t} \bm{C}_i  & \bm{0} \\
									     \bm{0} & \bm{\tilde{C}}_t                & \oplus_i \bm{X}_{i,t} \bm{C}_i               & \bm{0} \\
											 \bm{0} & \bm{0}                         & \bm{0}                                       & \bm{C}
							  \end{pmatrix}	
								\begin{pmatrix}																													
										\text{vec} ( \{ \imath \bm{v}^{(d)}_{i,t} \} ) \\
										\text{vec} ( \{ \bm{v}_{i,t} \} ) \\
										\text{vec} ( \{ v^{(n)}_{i} \} )										
								\end{pmatrix},
\end{array}
\label{eq:TVPSURGLLSPwindow}
\end{equation}
where the hyperbolic norm is used together with the imaginary unit $\imath$ to downdate the estimate of the TVP-SUR model \citep{rader:86,rader:88}. Namely, for a complex vector  \( \bm{x} \), the hyperbolic norm gives \( \left\| \bm{x} \right\|_h = \bm{x}^H \bm{\Psi} \bm{x} \) where \( \bm{\Psi} \) is a signature matrix and \( ( \cdot )^H \) denotes the conjugate transpose. 
Here, \( \oplus_i \imath \bm{x}^{(d)}_{i,t} \bm{C}_i  \) and \( \oplus_i \bm{X}_{i,t} \bm{C}_i \) is the new information incorporated into the variance covariance matrix of the first $t$ observations due to the inclusion of the new data point. Notice that the information which updates the covariance matrix of the deleted observations, i.e. \( \bm{x}^{(d)}_{i,t} \bm{C}_i \), is multiplied with the imaginary unit since it has to be excluded from the model. The GLLSP in \refeq{TVPSURGLLSPwindow} is solved by computing the corresponding RQ and generalised QR decompositions in a manner similar to the updating but using hyperbolic transformations when information needs to be removed from the TVP-SUR model \refeq{TVPSUR1}.

\section{Computational Experiments}\label{sec:Experiments}

\indent\par
Experiments have been designed to assess the computational efficiency of the proposed algorithms. Specifically, the strategies presented herein have been compared with existing ones which estimate the model afresh. 
The \textit{computational efficiency} of one algorithm compared to another algorithm is defined as the ratio of the computational cost of the two algorithms. Here, the execution time (in seconds) required by each algorithm to compute the desired estimate is presented in order to determine the \textit{computational efficiency} of the proposed strategies. 

To analyse the computational performance of the proposed methods and their counterparts, experiments based on synthetic data have been conducted. For the efficient implementation of the new methods, sequential and recursive strategies which exploit the special sparse structure of the matrices are employed  \citep{yanev:07,hadjiantoni:16,hadjiantoni:18}. Three cases with the corresponding algorithms have been considered. 
Specifically, it is assumed that the TVP-SUR model has been estimated using the initial dataset and then, new observations enter the dataset and/or old observations are deleted. 

Firstly, the problem of estimating the TVP model \refeq{MTVPadd1}, which incorporates the effect of a single observation is investigated, by estimating the model \textit{afresh} (see Algorithm~\refalg{AlgTVPSUR} in Section~\refsec{RecursiveTVPSUR}) 
and by implementing the new \textit{Updating} algorithm (see Algorithm~\refalg{AlgUTVPSUR} in Section~\refsec{UpdateTVPSURmodel}) which solves the GLLSP \refeq{UTVPSURGLLSP1}. Table~\reftab{TVPtable1} presents the execution times, in seconds, of both algorithms which recursively add the effect of one new observation into the model $100$ times. That is, the execution times presented in the third and fourth columns of Table~\reftab{TVPtable1} are the sum of re-estimating the model with one extra observation $100$ times. Examples with various numbers of time-varying regressions $G$ and unknown parameters $K$ are shown. The times in Table~\reftab{TVPtable1} confirm that the \textit{Updating} algorithm outperforms significantly the \textit{Afresh} algorithm for the estimation of the TVP model. The \textit{computational efficiency} of the proposed method increases as the dimensions of the models increase. 

\begin{table}[h!]
\centering
\caption{Execution times in seconds of the recursive estimation of the TVP-SUR model.}
\begin{tabular}{ccllc}
\addtolength{\tabcolsep}{-20pt}
$G$  & $K$    & Afresh  & \text{Updating}  & $\displaystyle \frac{\text{Afresh}}{\text{Updating}}$ \\ \hline\hline
25   & 100    & 5 		     & 1       		  & 6   \\ 
50   & 200    & 23  			 & 3    			  & 9   \\
75 	 & 300    & 63			   & 6 				    & 10 	\\ 
100  & 400    & 135			   & 12 			    & 12  \\  \hline
250  & 1000   & 2068 			 & 182  			  & 11  \\ 
500  & 2000   & 19194			 & 1420 			  & 14  \\ 
750  & 3000   & 73248			 & 4654  			  & 16  \\ 
1000 & 4000   & 198116		 & 11495			  & 17    \\ \hline\hline
\end{tabular}

\begin{minipage}[t]{0.9\columnwidth}
\small
The \textit{Afresh} and \textit{Updating} algorithms estimate the TVP-SUR model with one new observation at $s=100$ points in time. Models with different numbers of regressions $G$ and parameters $K$ are presented. 
\end{minipage} 
\label{tab:TVPtable1}
\end{table}

Secondly, consider deriving the smoothing estimates in \refeq{SmoothingTVPSUR} by solving the GLLSP \refeq{SmoothingTVPSURGLLSP1} \textit{afresh} or by solving GLLSP \refeq{GLLSPsmothing} which utilises previous computations. The \textit{Afresh} smoothing algorithm is a variation of Algorithm~\refalg{AlgTVPSUR} whereas the new \textit{Revising} algorithm is Algorithm~\refalg{AlgSmoothingTVPSUR} in Section~\refsec{SmoothingTVPSURmodel}. Table~\reftab{TVPtable2} compares the two algorithms when each of them, at the end of the period $M=60$, goes backwards $5$ points in time to compute the smoothing estimates. 
That is, \( \bm{\beta}_{M-i|M} \), \( i=1,\dots,5 \) are estimated. The time presented in each case is the average time required of the corresponding algorithm after $100$ iterations. The \textit{Revising} algorithm is considerably computationally more efficient than the algorithm that solves the GLLSP \refeq{SmoothingTVPSURGLLSP1} \textit{afresh}. The computing performance of the \textit{Revising} algorithm becomes more effective when both the number of regressions $G$ and the number of unknown parameters $K$ increase. 

\begin{table}[h!]
\centering
\caption{Execution times in seconds of the smoothing estimates for the TVP-SUR model.}
\begin{tabular}{ccllc}
\addtolength{\tabcolsep}{-20pt}
$G$  & $K$   & \text{Afresh}  & \text{Revising}  & $\displaystyle \frac{\text{Afresh}}{\text{Revising}}$ \\ 
     &       & $\times100$               &  $\times100$       & \\ 
\hline\hline
10 	 & 100   & 5       & 0    & 13   \\ 
25   & 100   & 50      & 1    & 45    \\
50   & 100   & 332     & 5    & 66    \\ \hline
10   & 500   & 6       & 5    & 1    \\ 
50    & 500   & 341     & 14   & 24    \\
100   & 500   & 2425    & 35   & 69    \\ \hline
 50   & 1000  & 344     & 55   & 6    \\
100  & 1000  & 2455    & 115  & 21    \\ 
250  & 1000  & 35467   & 184  & 193    \\ \hline
\end{tabular}
\begin{minipage}[t]{0.9\columnwidth}
\small
The \textit{Afresh} and \textit{Revising} algorithms go backwards at $s=5$ points in time to compute the smoothing estimates of the TVP-SUR model. Models with $K=100,500,1000$ unknown parameters and different number of regressions are estimated. 
The execution times presented is the overall time required to look backwards at $5$ points in time. The average times of $100$ such repetitions, multiplied by 100, are reported. 
\end{minipage} 
\label{tab:TVPtable2}
\end{table}

Finally, consider estimating the model over a rolling window of data. Namely, let the fixed size estimation window move forward at one point of time to capture the information from the next data point and while excluding the effect of the oldest data point. 
That is, estimate \refeq{TVPSURwindow} by employing Algorithm~\refalg{AlgTVPSUR} or by solving the GLLSP \refeq{TVPSURGLLSPwindow} using the \textit{up-downdating} algorithm in \citep{hadjiantoni:16}. Table~\reftab{TVPtable3} reports the total time to estimate the model over a window (of fixed size) which rolls ahead one data point $100$ times. The ratios of the execution times in Table~\reftab{TVPtable3} confirm that the recursive \textit{Up-downdating} algorithm performs better than the \textit{Afresh} algorithm and, similarly to the previous computational results, the efficiency increases when the models' dimensions increase. 

\begin{table}[h!]
\centering
\caption{Execution times in seconds for the window estimation of the TVP-SUR model.}
\begin{tabular}{ccllc}
\addtolength{\tabcolsep}{-20pt}
$G$ & $K$    & \text{Afresh}  & \text{Up-downdating}  & $\displaystyle \frac{\text{Afresh}}{\text{Up-downdating}}$ \\ \hline\hline
10   & 250   & 8     & 6   & 1   \\
25   & 250   & 43    & 8   & 6   \\
50   & 500   & 238   & 36  & 7    \\
100  & 500   & 1059  & 95  & 11    \\ 
250  & 1000  & 12516 & 800 & 16    \\
\end{tabular}

\begin{minipage}[t]{0.9\columnwidth}
\small
The \textit{Afresh} and \textit{Up-downdating} algorithms estimate the TVP-SUR model over a rolling window of data where one observation is added to the model and one is deleted. Models with initial number of observations $t=59$, $G=10,25,50,100,250$ regressions and different numbers of parameters $K$ are presented. The time required to up-downdate the model with one observation $100$ times is presented. 

\end{minipage} 
\label{tab:TVPtable3}
\end{table}

Overall, the results show that the recursive algorithms which utilise previous computations outperform the algorithms which estimate the model afresh. 
The computational efficiency becomes notable when both $G$ and $K$ increase. 
Experiments have been conducted for models of other dimensions; the computational efficiency is similar to Tables~\reftab{TVPtable1},~\reftab{TVPtable2} and \reftab{TVPtable3}. 
The results show the practical usability of the proposed methods in estimating TVP models of high dimensions.
 
\section{Conclusions and Future Work}\label{sec:Conclusions}

\indent\par
The estimation of the multivariate TVP model using alternative numerical methods has been investigated. The TVP model can be written as a general linear model and therefore be estimated with the method of GLS \citep{sant:77}. However, using GLS to estimate such a model is computationally expensive and numerically inaccurate due to the computation of large matrix inverses. Therefore, the proposed method considers the equivalent GLLSP to provide the estimates of the model. The GLLSP method has been shown to be computationally faster and numerically more stable than solving the normal equations to obtain the GLS estimator \citep{paige:79}. 

Herein, numerical methods have been investigated for the efficient estimation of the TVP and TVP-SUR models. Various cases have been examined for the efficient estimation of the model when the estimates of the unknown parameters need to be re-computed after changes occur in the dataset. Specifically, the case of updating the model with one new observation and also that of deriving the smoothing estimates of the model are examined. Finally, the simultaneous addition and deletion of observations (up-downdating) within the context of rolling window estimation is explored. 
 
The algorithms developed herein, take advantage of the special sparse structure of the models and utilise efficiently previous computations. Experiments have been carried out to analyse the computational performance of the proposed algorithms which update the model with one new observation, compute the smoothing estimates of the model and estimate the model over a rolling window of data. The computational results show that the proposed algorithms are computationally more efficient than their counterparts and that their performance becomes more significant in high dimensions. This demonstrates the usefulness of the proposed methods in practical problems of large-scale TVP models.

Future work will consider the estimation of multivariate TVP models using a high-dimensional setting where the number of covariates exceeds the sample size, resulting in a singular variance covariance matrix. Having more parameters to estimate than available observations will affect the estimation of the initial model (see \refeq{TVPSUR1}), but the updating when an extra data point arrives will be straightforward using the novel methods developed herein. Furthermore, models with a more complex time-varying structure should be addressed. For example, to allow for a time-varying variance-covariance matrix. Additionally, the estimation of time-varying parameter vector autoregressive models using the proposed numerical methods and their extension to model selection merit investigation.  

\singlespacing
\addcontentsline{toc}{section}{Bibliography}
\bibliographystyle{apalike}
\bibliography{TimeVaryingParametersModel_bibliography}

\end{document}